\let\myover=\over
\documentstyle[a4,12pt,amsmath,amssymb,epsfig,amsfonts]{article}

\def\e{{\rm e}}
\def\al{\alpha}
\def\bt{\beta}

\def\d{\partial}
\def\l{\left(}
\def\r{\right)}

\newcommand{\be}{\begin{equation}}
\newcommand{\ee}{\end{equation}}

\newcommand{\Tr}{{\rm Tr}}
\newcommand{\bg}{\begin{gather}}
\newcommand{\eg}{\end{gather}}

\let\over=\myover
\begin{document}
\begin{center}
  {\Large\bf Quasi-localized states on
noncommutative solitons } \\
  \medskip S.~L.~Dubovsky, V.~A.~Rubakov, S.~M.~Sibiryakov\\
    \medskip
  {\small
     Institute for Nuclear Research of
         the Russian Academy of Sciences,\\  60th October Anniversary
  Prospect, 7a, 117312 Moscow, Russia\\
  }  
\end{center}
\begin{abstract}
We consider noncommutative gauge theories which have zero mass
states propagating along both commutative and noncommutative
dimensions.  Solitons in these theories generically carry $U(m)$
gauge group on their world-volume.  From the point of view of
string theory, these solitons correspond to
 ``branes within branes''. We show that once the world-volume
$U(m)$ gauge theory is in the Higgs phase, light states become
quasi-localized, rather than strictly localized on the soliton,
{\it i.e.} they mix with light bulk modes and have  finite
widths to escape into the noncommutative dimensions. At small
values of $U(m)$ symmetry breaking parameters, these widths are
small compared to the corresponding masses. Explicit examples
considered are adjoint scalar field in the background of a
noncommutative vortex in $U(1)$-Higgs theory, and gauge fields
in instanton backgrounds in pure gauge noncommutative
theories.  
\end{abstract}

\section{Introduction and summary}

Quasi-localization, rather than perfect localization, of states on
a brane is a common property of various brane-world 
models
\cite{Charmousis:1999rg,Kogan:1999wc,Dvali:2000hr,Dvali:2000rx,Dubovsky:2000am,Dubovsky:2001fj}.
Particles may not be trapped to a brane forever, but may have
finite, albeit small, probability to escape into extra dimensions.
This phenomenon may occur even at low energies,
provided bulk modes have continuum spectrum starting from
zero energy and there is mixing between brane modes and 
continuum modes (for a review see, e.g., 
Ref. \cite{Rubakov:2001kp}). Clearly,
this possibility is of interest for
phenomenology, and also for
the study of the properties of branes in a more
theoretical context.

Particularly interesting field theory branes are
noncommutative solitons (for reviews see 
Refs. \cite{Nekrasov:2000ih,Harvey:2001yn,Konechny:2001wz}).
Noncommutative field theory arises in an
appropriate limit of string 
theory \cite{Connes:1997cr}, and the properties of many
noncommutative solitons match nicely the properties
of $D$-branes. Indeed, it has been 
suggested \cite{Pilo:2000nc} to use the noncommutative
solitons for constructing phenomenologically 
acceptable brane-world models.

In this paper we discuss quasi-localization of states on solitons in
noncommutative gauge theories. The class of models we consider is the
one where the gauge (and possibly matter) fields have continuum of
bulk modes, weakly coupled at low energies\footnote{We live aside unstable
solitons corresponding to $D$-branes in tachyonic vacuum
\cite{Harvey:2000jt,Gopakumar:2000rw}.}. 
From the point of view of string theory,
solitons in these models correspond to ``branes within branes'',
see, e.g., Ref.
\cite{Douglas:1995bn} and references therein.

Generically, a soliton of this type has $U(m)$ gauge symmetry on its
world-volume. If (part of) this gauge symmetry is unbroken, 
charged matter fields as well as massless gauge fields are strictly
localized on the soliton. Our main observation is that the situation
changes if $U(m)$ on the soliton world-volume is in the Higgs phase:
gauge and/or matter fields become quasi-localized. For fields that
have massless bulk modes, escape into extra dimensions occurs even at
low energies; at small value of the parameter of $U(m)$ symmetry
breaking, the life-times against this escape are large compared to the
inverse masses of the quasi-localized modes.

To introduce the mechanism of quasi-localization most explicitly, we
consider in Sect.2 a simple example of a $m$-vortex solution in $U(1)$
gauge-Higgs theory in two noncommutative and $p$ commutative spatial
dimensions 
\cite{Polychronakos:2000zm,Jatkar:2000ei,Bak:2000ac,Bak:2000im}. 
The bulk modes of both gauge and Higgs fields are
massive in this model, so escape of these fields into the
noncommutative dimensions does not occur at low energies. We
introduce an extra adjoint scalar field which has massless modes in
the bulk, and show that its states become quasi-localized on the vortex
even at low energies, provided that $U(m)$ gauge theory on the vortex
is in the Higgs phase. We calculate the widths of the quasi-localized
states at small values of the parameter governing $U(m)$ symmetry
breaking, and find that these widths are parametrically
 smaller than the masses of
these states. In this model, as well as in another example studied in
this paper, there is a hierarchy of life-times of different
quasi-localized modes. In the case of vortex, this hierarchy is
related to the rotational symmetry of the background: 
we will see that
higher angular momentum modes live {\em longer} on the soliton,
because certain mixing terms of these modes are either forbidden by
rotational symmetry or suppressed due to the centrifugal barrier. 

In Sect.3 we discuss quasi-localization on noncommutative solitons in
more general terms. We develop perturbation theory in $U(m)$ symmetry
breaking parameters  and analyze what kind of mixing between brane
modes and bulk modes appears to the first and second orders. 

In Sect.4 we study pure gauge theory in four noncommutative
and $p$ commutative spatial dimensions. Instantons in this theory
\cite{Nekrasov:1998ss,Aganagic:2000mh,Furuuchi:2000vc,
Harvey:2000jb,Furuuchi:2000dx} 
correspond to $Dp-D(p+4)$ system. ``Zero-size'' anti-self-dual 
$m$-instantons in a
theory with anti-self-dual noncommutativity 
\cite{Aganagic:2000mh,Harvey:2000jb} (which are actually
non-singular solutions)
have unbroken, strictly localized $U(m)$ gauge theory on their
world-volume. We show that once the instanton size is non-vanishing,
the gauge theory on the instanton world-volume not only is in the Higgs
phase, but also becomes quasi-localized even at low energies. We
consider explicitly the case of  small instanton size, which
corresponds to small gauge boson masses on the soliton, and show that
the widths of the quasi-localized gauge bosons against the escape into
noncommutative dimensions are again small compared to the masses of
these states. Whether or not these widths exhibit the hierarchical
pattern depends on the structure of the instanton: the hierarchy is
absent in examples considered in Sect.4, while in another example
sketched in Appendix the widths are hierarchical.

It appears that the quasi-localization of low-energy theory on
noncommutative solitons is generic in models having massless modes in
the bulk and gauge theories in the Higgs phase on the soliton
world-volume. It is tempting to speculate that in string theory
motivated brane-world models, massive particles which are neutral
under electric charge and color may be unstable against escape into
extra dimensions. On the other hand, in the context of noncommutative
theories without gravity, particles carrying unbroken charges of the
brane-world gauge theory, and also massless gauge bosons of that theory
are trapped to the solitonic brane forever (in other words, processes
like $e^-\to nothing$ or $\gamma \to nothing$
are not allowed, unlike in some other
brane-world models \cite{Dubovsky:2000av,Dvali:2000rx,Dubovsky:2001fj}). 
It remains to be 
understood whether
or not this property still holds when gravity is turned on.

\section{Quasi-localization on  noncommutative vortex}
\label{vortex}
\subsection{Vortex solutions}
In this section we consider an $U(1)$ gauge
theory with fundamental Higgs field in $(1_{\mathrm
time}+p+2)$-dimensional
space-time with two space-like noncommutative dimensions $x_1,\;
x_2$. 
The action for this theory has the following form,
\begin{equation}
\label{action}
S={1\over g^2}\int d^{p+1}y\;d^2x\left[-{1\over 4} F_{AB}*F^{AB}+
D_A\phi^+*D^A\phi-{1\over 2}(\phi^+*\phi-v^2)^2\right]
\end{equation}
where $y^{\mu}$ are commuting dimensions,
\begin{eqnarray}
F_{AB}&=&\d_AA_B-\d_BA_A-i(A_A*A_B-A_B*A_A)\nonumber\\
D_A\phi&=&\d_A\phi-iA_A*\phi
\end{eqnarray}
and the Moyal product is defined as follows,
\[
f(x)*g(x)\equiv\left.\e^{-i{\theta\over
2}\epsilon^{ij}\d_i\d_j'}f(x)g(x')
\right|_{x=x'}\;.
\]
As shown in
Ref.~\cite{Polychronakos:2000zm,Jatkar:2000ei,Bak:2000ac,Bak:2000im}, 
this theory admits static
soliton solutions independent of the commuting coordinates and
having a form of a vortex in the noncommutative plane $x_1,x_2$. 
To describe these solitons, let us switch to the Fock space
notations. Then the energy density takes the following
form,
\[
E={2\pi\theta\over g^2}\Tr\left\{ {1\over
2\theta^2}([C,C^+]+1)^2+D_z\phi^+ D_{\bar{z}}\phi+D_{\bar{z}}
\phi^+ D_{z}\phi+{1\over 2}(\phi^+\phi-v^2)^2\right\}\;,
\]
where
\begin{gather}
z=\sqrt{1\over 2}(x_1+ix_2)~,~~[z,\bar{z}]=\theta\nonumber\\
C=a^++i\sqrt{\theta}A_z\nonumber\\
a={z\over\sqrt{\theta}},a^+={\bar{z}\over\sqrt{\theta}}\;\nonumber\\
D_z\phi=-{1\over\sqrt{\theta}}[a^+,\phi]-iA_z\phi\nonumber\\
D_{\bar{z}}\phi={1\over\sqrt{\theta}}[a,\phi]-iA_{\bar{z}}\phi
\nonumber
\end{gather}
The properties of stable vortex solutions depend on the
value of the parameter $\theta v^2$. At $\theta v^2\geq 1$ stable
vortex is non-BPS and can be obtained by the solution generation 
technique~\cite{Bak:2000ac,Bak:2000im,Harvey:2000jb,Hamanaka}. 
The $m$-soliton solution has the following form in this
case (for  all solitons located at one point in the noncommutative plane)
\begin{eqnarray}
\label{technique}
\phi&=&v S_m^+\nonumber\\
C^+&=&S_m^+aS_m\;
\end{eqnarray}
where $S_m^+$ is the shift operator,
\[
S_m^+=\sum_{n=0}^{\infty}|n+m\rangle\langle n|\;.
\]

In other words, the matrices of the operators $C^+$ and $\phi$ in the
Fock basis have the form 
\begin{gather}
\label{*}
\begin{array}{l}
~~~~~~~~~~~\overset{\displaystyle m}{\overbrace{~~~~~~~~~~~~~~~~}} \\
C^+=\begin{pmatrix}
0&0&\dots&0&&0&0&\dots \\
\hdotsfor{4}&&\hdotsfor{3}\\
0&0&\dots&0&&0&0&\dots \\
0&0&\dots&0&&&\mbox{\vbox to 0pt{\hbox{\Huge $a$}}}& \\
\hdotsfor{4}&&&&\\
0&0&\dots&0&&&& 
\end{pmatrix} 
\begin{array}{c} 
\left. \begin{array}{c}\\ \\ \\ \end{array}\right\}m 
\\ \\ \\ \\ \end{array}
\end{array}
\\ \nonumber \\
\phi=
\begin{pmatrix}
0&0&0&\dots \\
\hdotsfor{4}\\
0&0&0&\dots \\
v&0&0&\dots \\
0&v&0&\dots \\
0&0&v&\dots \\
\hdotsfor{4}
\end{pmatrix}
\begin{array}{c} 
\left. \begin{array}{c}\\ \\ \\ \end{array}\right\}m
\\ \\ \\ \\ \\ \end{array}
\end{gather}
When considered as a $p$-brane in $(p+2)$-dimensional space, this
soliton carries on its world-volume 
localized gauge fields  corresponding to the
unbroken $U(m)$ subgroup of the original $U(\infty)$ group. In the
Fock basis, the gauge fields of $U(m)$ are $m\times m$ matrices in the
upper left corner (cf. Eq.~(\ref{*})). The corresponding wave
functions in the coordinate representation are localized near the
vortex and have transverse size of order $\sqrt{\theta}$.
There is also  continuous spectrum of gauge 
fields 
corresponding to the broken generators of $U(\infty)$. 
The latter fields propagate in the bulk.
Non-zero vacuum
expectation value of the field $\phi$ at $|z|\to \infty$ (see
Eq.~(\ref{technique})) provides a mass gap $gv$ to this continuum.

At $\theta v^2<1$ the solution given by Eq.~(\ref{technique}) is
unstable. Instead, there exists a BPS solution of the following form
~\cite{Bak:2000im}
\begin{eqnarray}
\label{BPS}
\phi&=&v\sum_{n=0}^{\infty}(1+\phi_n)|n+m\rangle\langle n|\nonumber\\
C^+&=&\mu_m+\epsilon_m+S_m^+\tilde{a}S_m\;,
\end{eqnarray}
where 
\begin{gather}
\mu_m\equiv 
\sum_{\alpha=1}^{m-1}\sqrt{\alpha(1-\theta
v^2)}|\alpha-1\rangle\langle 
\alpha|\;,
\label{muvortex}
\\
\epsilon_m\equiv \sqrt{m(1-\theta v^2)}|m-1\rangle\langle m|\;,
\label{epsilonvortex}
\end{gather}
and
\begin{equation}
\label{R*}
\tilde{a}\equiv\sum_{n=1}^{\infty}(\sqrt{n}+c_n)|n-1\rangle\langle n|\;.
\end{equation}
The corresponding matrices in the Fock basis are
\begin{gather}
\label{C*}
\rule{0pt}{5pt}
\begin{array}{l}
~~~~~~~~~~\overset{\displaystyle m}{\overbrace{~~~~~~~~~~~~~~~~~~~~~~~~~~~~~~~~~~~~~~~}} \\
C^+=\begin{pmatrix}
0&\sqrt{\omega}&0&\dots&0&&0&0&\dots \\
0&0&\sqrt{2\omega}&\dots&0&&0&0&\dots \\
\hdotsfor{5}&&\hdotsfor{3}\\
0&0&0&\dots&\sqrt{(m-1)\omega}&&0&0&\dots \\
0&0&0&\dots&0&&\sqrt{m\omega}&0&\dots \\
0&0&0&\dots&0&&&\mbox{\vbox to 0pt{\hbox{\Huge $\tilde{a}$}}}& \\
\hdotsfor{5}&&&&\\
0&0&0&\dots&0&&&& 
\end{pmatrix} 
\begin{array}{c} 
\left. \begin{array}{c}\\ \\ \\ \\ \\ \end{array}\right\}m 
\\ \\ \\ \\  \end{array}
\end{array}
\\ \nonumber \\
\phi=
\begin{pmatrix}
0&0&0&\dots \\
\hdotsfor{4}\\
0&0&0&\dots \\
v_1&0&0&\dots \\
0&v_2&0&\dots \\
0&0&v_3&\dots \\
\hdotsfor{4}
\end{pmatrix}
\begin{array}{c} 
\left. \begin{array}{c}\\ \\ \\ \end{array}\right\}m
\\ \\ \\ \\ \\ \end{array}
\end{gather}
where
\[
\omega\equiv\sqrt{1-\theta v^2}
\]
and $v_n=v(1+\phi_n)$.
The coefficients $\phi_n$ and $c_n$ are determined by a set of recursion
relations which were obtained in Ref.~\cite{Bak:2000im}. In what follows we
consider the case when the parameter $\omega$
is small, 
\[
\omega\ll 1\;.
\] 
In this case $\phi_n$ and $c_n$ are also small and their
explicit form is not essential for our purposes. The relevant property
of these coefficients is that they tend to zero as $n$ tends to
infinity,
\begin{equation}
\label{zerolimit}
\lim_{n\to\infty} \phi_n,c_n=0\;.
\end{equation}
$\mu_m$ and $\epsilon_m$ serve as vacuum expectation values
 of the adjoint and fundamental
Higgs fields giving masses
to the gauge bosons of the $U(m)$ gauge group on the vortex. 
As a result, this gauge group is 
spontaneously broken completely. 
In addition, $\epsilon_m$ introduces mixing between $U(m)$ gauge
bosons and 
gauge
bosons from the continuum spectrum. However, the latter has a mass
 gap, the gauge bosons from continuum can be
integrated out at low energies, and this mixing does not lead
to any interesting low energy effects at $\omega\ll 1$.

\subsection{Adjoint scalar}
Let us now introduce additional massless
real adjoint scalar field $f$ with the action
\begin{equation}
\label{saction}
{2\pi\theta\over g^2}\int
d^{p+1}y\;\Tr\left\{-{1\over\theta}[C,f][C^+,f]+D_{\mu}f D_{\mu} f\right\}\;,
\end{equation}
where
\[
D_{\mu}f=\d_{\mu}f-i[A_{\mu},f]\;.
\]
Let us study  the mass spectrum of the field $f$ in the vortex
background. A non-trivial mass matrix for $f$ is provided by the first
term in the action (\ref{saction}). It is convenient to decompose the
 field $f$ in the following way,
\begin{equation}
\label{D*}
\begin{array}{l}
~~~~~~~~\;\overset{m}{\overbrace{}} \\
f=\left(\begin{array}{cc}
 \psi & \xi \\ \xi^+ & \chi
\end{array}\right) \begin{array}{c} \}m\\ \vphantom{d} \end{array}
\end{array}
\end{equation}                                                                                       
In other words,
\[
f=\psi+\xi+\xi^++\chi\;,
\]
where $\psi$ is a Hermitean $m\times m$ matrix
\[
\psi=P_mfP_m\equiv\sum_{\alpha,\beta=0}^{m-1}\psi_{\al}^{\bt}
|\al\rangle\langle \bt|\;,
\]
$\xi$ is a $m\times\infty$ matrix
\[
\xi=P_mf(1-P_m)\equiv\sum_{\al=0}^{m-1}\sum_{n=0}
^{\infty}\xi_{\al}^n|\al\rangle\langle m+n|\;,
\]
and $\chi$ is a Hermitean $\infty\times\infty$ matrix
\[
\chi=(1-P_m)f(1-P_m)\equiv\sum_{k,n=0}^{\infty}\chi_k^n|m+k\rangle\langle
m+n|\;.
\]
Here $P_m$ is a projector
\[
P_m=\sum_{\al=0}^{m-1}|\al\rangle\langle \al|\;.
\]
In the case of non-BPS vortex (\ref{technique}) with unbroken $U(m)$
 gauge group on
 its world-volume, it is straightforward to
check that the fields $\psi_{\al}^{\bt}(y)$ are massless scalar fields
 strictly
localized on the vortex world-volume and belonging to adjoint
representation of $U(m)$, $\xi_{\al}^n(y)$ are massive fundamentals with
masses
\begin{equation}
\label{heavymass}
m_n^2={2n+1\over\theta}
\end{equation}
and $\chi_{k}^n(y)$ are fields in the gapless continuum spectrum, which
are neutral under $U(m)$.
 
\subsection{Quasi-localization on single vortex}
Let us first discuss the spectrum of the adjoint scalar field
at small but non-zero $\omega$ in the
background of one-vortex solution, $m=1$.
In the matrix notations (\ref{C*}) and (\ref{D*}),
the first term in the action
(\ref{saction}) takes the following form,
\begin{equation}
\label{U1}
\begin{split}
-{1\over\theta}\Tr\left\{[C,f][C^+,f]\right\}
=&-{1\over\theta}\Bigl[
\Tr\left\{[\tilde{a},\chi][\tilde{a}^+,\chi]\right\}
\\ 
&+\sum_{n=1}^{\infty}(2\tilde{n}_1-1)\xi^{n-1}\xi_{n-1}^*+
\sum_{n=1}^{\infty}\omega^2\chi_{0}^n
\chi_{n}^{0}
+\omega^2\chi_{0}^0\chi_{0}^0+\omega^2\psi^2
\\ 
&-\sum_{n=1}^{\infty}2\omega
\sqrt{\tilde{n}_2}(\chi^{0}_n\xi^{n-1}+\chi_{0}^n\xi_{n-1}^*)
-2\omega^2\chi_{0}^0\psi\Bigr]
\end{split}
\end{equation}
where
\[
\tilde{n}_{1,2}=n+O(\omega)\;.
\]
The first term in Eq.~(\ref{U1}) describes the propagation of the
field $\chi$ along noncommutative dimensions: with appropriate
redefinition of noncommutative coordinates it becomes merely the
gradient term,
$\int d^2z'\d_{z'}\chi\d_{\bar{z}'}\chi+O(\omega)$. Due to the
property (\ref{zerolimit}), the propagation is free far away from the
vortex, and the spectrum still starts from zero energy. The second
term in Eq.~(\ref{U1}) gives large masses (\ref{heavymass}) (up to
small corrections) to the fundamentals $\xi^n$. The third, fourth and
fifth terms provide small diagonal masses to the fields $\chi^n_0$,
$\chi^0_0$ and $\psi$, respectively.
Most importantly, there is a term that mixes the would-be localized
state $\psi$ with the states from continuum: this is the last term in
Eq.~(\ref{U1}). 
Integrating out massive fields $\xi^n$ one arrives at the following
effective mass terms for light fields $\chi_n^k$ and $\psi$
(neglecting $O(\omega)$ corrections in $\tilde{n}_{1,2}$)
\begin{equation}
\begin{split}
{\it M}=&-{1\over\theta}\Tr\left\{[\tilde{a},\chi][\tilde{a}^+,\chi]\right\}+
\sum_{n=0}^{\infty}
{\omega^2\over\theta}
\l1-{4n\over 2n-1}\r \chi_{0}^{n*}\chi_{0}^{n}
\\
&-2{\omega^2\over\theta}\chi_{0}^{0}
\psi+{\omega^2\over\theta}\psi^2\;.
\end{split}
\label{light}
\end{equation}
This effective mass terms are relevant at energy scales below
$1/\sqrt{\theta}$. One observes that at small $\omega$, there is an
interesting low energy scale $\omega/\sqrt{\theta}$, so we concentrate
on physics at this scale. 

We begin with the first two terms in Eq.~(\ref{light}). They
contain corrections to the quadratic action of the fields $\chi_0^n$,
which at first glance appear relevant at the scale
$\omega/\sqrt{\theta}$.
 Let us see that
this is not the case.

Let us come back to coordinate formulation of the noncommutative
theory, and write the field $\chi(y,z,\bar{z})$ in Fourier
representation along noncommutative dimensions,
\begin{equation}
\chi(y,z,\bar{z})=\int {d^2k\over (2\pi)^2}
\tilde{\chi}_k(y)e^{i(k_zz
+k_{\bar{z}}\bar{z})}\;.
\label{fourier}
\end{equation}
We are interested in low-momentum part, $k\sim
\omega/\sqrt{\theta}$. The components $\chi_0^n$ entering the second
term in Eq.~(\ref{light}) are
\begin{equation}
\chi_0^n={i^n\over\sqrt{n!}}\int {d^2k\over (2\pi)^2}\e^{-k^2\theta/2}
\tilde{\chi}_k(y)(k_{\bar{z}}\sqrt{\theta})^n
~\approx~ {i^n\over\sqrt{n!}}\int {d^2k\over (2\pi)^2}
\tilde{\chi}_k(y)(k_{\bar{z}}\sqrt{\theta})^n
\label{fourier1}
\end{equation}
Since $d^2k\propto \omega^2/\theta$, the second term in
Eq.~(\ref{light}) is at least of order $(\omega^6/\theta^3 \cdot
|\tilde{\chi}_k|^2)$, whereas the gradient term is of order 
$${1\over\theta}\int d^2 k~ k^2 |\tilde{\chi}_k|^2 \sim \omega^4/\theta^3\cdot
|\tilde{\chi}_k|^2\;.$$ 
Thus, the second term in Eq.~(\ref{light}) is
small at small $\omega$ and $k\sim \omega/\sqrt{\theta}$. Similar
argument applies to the corrections to the gradient term coming from
the fact that $\tilde{a}$ and $a$ differ by $O(\omega^2)$.

Neglecting the higher-order terms, we write the low energy effective
action as follows,
\begin{equation}
\begin{split}
S_{eff}={2\pi\theta
\over g^2}\int d^{p+1}y & \l (\d_{\mu}\psi)^2-{\omega^2\over\theta}\psi^2+\right.\\
&\left. +\int {d^2k\over 2\pi\theta(2\pi)^2}\l
\d_{\mu}\tilde{\chi}_k\d_{\mu}\tilde{\chi}_{-k}-k^2\tilde{\chi}_k
\tilde{\chi}_{-k}+4\pi\omega^2\tilde{\chi}_k\psi
\r\r.
\end{split}
\label{effective}
\end{equation}
From this action one obtains the following equations determining
the propagator $G_{\psi}(p)$ of the field $\psi$,
\begin{eqnarray}
\label{first}
\l p^2-{\omega^2\over\theta}\r G_\psi(p)+{\omega^2\over\theta}\int{d^2k\over
(2\pi)^2}\tilde{\chi}_k&=&1\\
\label{second}
\l p^2-k^2\r\tilde{\chi}_k
+2\pi\omega^2G_{\psi}(p)&=&0
\end{eqnarray}
where $p^{\mu}$ is the momentum along commutative dimensions.
Expressing $\tilde{\chi}_k$ from Eq.~(\ref{second}) and 
substituting it into Eq.~(\ref{first}) one obtains that the
propagator $G_{\psi}(p)$ has the Breit--Wigner form
\[
G_{\psi}(p)={1\over p^2-m^2+im\Gamma}
\]
with mass
\[
m^2={\omega^2\over\theta}
\]
and width
\begin{gather}
\label{Gvortex}
\Gamma= -{\omega^3\over 2\pi\sqrt{\theta}}{\mathrm{Im}}\int{d^2k\over
 p^2-k^2+i\epsilon}={\pi\omega^3\over 2\sqrt{\theta}}\ll m
\end{gather}
Thus, mixing between the field $\psi$ and fields $\tilde{\chi}_k$
from the continuum spectrum 
results in the delocalization of the field $\psi$. This field no longer
describes a
stable state localized on the vortex. Rather it
corresponds to a metastable resonance embedded in the continuum
spectrum. This state has a small but non-vanishing probability to escape 
from the brane.

\subsection{Multi-vortex case: hierarchy of widths}
Let us now consider the multi-vortex solution
with $m>1$. We still study the case $\omega\ll 1$, and
physics at energy scale $\omega/\sqrt{\theta}$. To understand 
what happens with field $f$ in this
case, it is convenient to make use of the symmetry under rotations in
the noncommutative plane, which is present in the action
(\ref{saction}) when the background field $C^+$ is given by
Eq.~(\ref{BPS}). 
Namely, this action is invariant under the
transformations
\[
f\to\e^{-i\alpha a^+a}f\e^{i\alpha a^+a}\;,
\]
leaving field $C$ invariant.
Explicitly, this rotation acts on the matrix elements of the operator
$f$ as follows,
\[
f_m^n\to\e^{i\alpha(n-m)}f_m^n
\]
where $f_m^n$ are defined by
\[
f=\sum_{m,n=0}^{\infty}f_m^n|m\rangle\langle n|\;.
\]
In other words, the field $f_m^n(y)$ has charge $(n-m)$ under this
symmetry. Clearly, this charge can be
interpreted as the angular momentum in the
noncommutative plane. Consequently, the action (\ref{saction}) in the
background field (\ref{BPS}) is the sum of the actions for fields with
different angular momenta. The fields with angular momentum $l$
combine into the matrix
\begin{gather}
\rule{0pt}{5pt}
\begin{array}{l}
~~~~~~~~~\overset{\displaystyle l}{\overbrace{~~~~~~~~~~~~~~~~~}} \\
f=\begin{pmatrix}
0&\dots&0&f_0^l&0&\hdotsfor{3} \\
0&\hdotsfor{2}&0&f_1^{(l+1)}&0&\hdotsfor{2} \\
0&\hdotsfor{3}&0&f_2^{(l+2)}&0&\dots\\
\hdotsfor{8}
\end{pmatrix}+h.c. 
\end{array}
\end{gather}
Due to the rotational symmetry, the fields with different $l$
 decouple.
Furthermore mixing between the states occurs between neighboring
entries of this matrix.

In terms of the fields $\psi$, $\xi$ and $\chi$ introduced in
Eq.~(\ref{D*}), the latter property implies that the would-be bound
states $\psi$ with non-zero angular momentum do not mix directly to
the continuum states $\chi$. Indeed, non-trivial mixing occurs between
the neighboring entries of the matrix (omitting indices of the
fields),
\begin{gather}
\rule{0pt}{5pt}
\begin{array}{l}
~~~~~~~~~~~~~~~~~~~~
\overset{\displaystyle l}{\overbrace{~~~~~~~~~~~~~~~~~}} 
 ~~\overset{\displaystyle m-l}{\overbrace{~~~~~~~~}}\\
\begin{array}{r} 
m-l\left\{ \begin{array}{c}\\ \\ \\ \end{array}\right. \\
l\left\{ \begin{array}{c} \\ \\ \\ \end{array}\right. 
\\ \\ \\ \\ \end{array}
\begin{pmatrix}
0&\dots&0&\psi&0&\dots&\hdotsfor{7} \\
\hdotsfor{6}&\hdotsfor{7}\\
0&\hdotsfor{3}&0&\psi&0&\hdotsfor{6}\\
0&\hdotsfor{4}&0&\xi&0&\hdotsfor{5}\\
\hdotsfor{6}&\hdotsfor{7}\\
0&\hdotsfor{5}&\dots&0&\xi&0&\hdotsfor{3}\\ 
0&\hdotsfor{5}&\hdotsfor{2}&0&\chi&0&\hdotsfor{2}\\
0&\hdotsfor{5}&\hdotsfor{3}&0&\chi&0&\dots\\
\hdotsfor{6}&\hdotsfor{7}
\end{pmatrix} +h.c.
\end{array}
\end{gather}
Thus, the fields $\psi$ mix between themselves and with heavy fields
$\xi$ only, and the decay of $\psi$ into continuum states occurs only
through weak mixing of the heavy fields $\xi$ between themselves and
finally with the fields $\chi$ propagating along noncommutative
dimensions. Clearly, this introduces the suppression of the decay
widths of the quasi-localized states $\psi$ with angular momentum $l$
by extra factor $\omega^{2l}$. Yet another suppression occurs due to
the fact that at low momenta, the components $\chi_m^n$ with
$(n-m)=l\neq 0$ are small, essentially due to the centrifugal barrier.  

Explicitly, the mass term for fields with angular momentum $l$
has the following
structure,
\begin{equation}
\begin{split}
\theta M_l=&-\Tr [a,\chi][a^+,\chi]-\Tr [\epsilon_m,\psi][\epsilon_m^+,\psi]+
\sum_{n=0}^{l-1}m_n^2|\xi_{m+n-l}^n|^2
\\
&+ \sum_{n=0}^{l-2}\omega\alpha_n\xi_{m+n-l}^n\xi_{m+n+1-l}^{n+1*}+
\omega^2\beta\psi_{m-l-1}^{m-1}\xi^{0*}_{m-l}+
\omega\gamma\xi_{m-1}^{l-1}\chi_{0}^{l*}+h.c.
\end{split}
\label{diagonal}
\end{equation}
where $\epsilon_m$ is defined in Eq.~(\ref{epsilonvortex})
 and is of order $\omega$,
the masses $m_n$ are the same as in Eq.~(\ref{heavymass}) modulo
small corrections proportional to $\omega$, coefficients
$\alpha_n$, $\beta$ and $\gamma$ are numbers of order one and we 
neglected $O(\omega^2)$-corrections involving the components of $\chi$ only.
Again, the first term in Eq.~(\ref{diagonal}) provides the kinetic
term along the noncommutative plane
for the field $\chi$.
The second term provides masses and mixings of
order $\omega^2$ for
quasi-localized fields $\psi_{\al}^{\bt}$. The third term makes the fields
$\xi_{\al}^n$ heavy. The terms on the second line lead to mixing
between quasi-localized modes $\psi$ and heavy fields $\xi$, as well
as to mixing between the heavy fields and states from continuum. We
see, that at
$l>0$ there is no direct mixing between $\psi$ and $\chi$. 
After integrating out the heavy fields, one
arrives at the following effective action, describing mixing between
the components of $\psi$ and $\chi$ of angular momentum $l$,
\begin{gather}
S_{eff}=\sum_i {2\pi\theta
\over g^2}\int d^4y\l |\d_{\mu}\psi_i|^2-\delta_i{\omega^2\over\theta}
|\psi_i|^2\nonumber\right.\\\left.
+\int {d^2k\over 2\pi\theta(2\pi)^2}\l
\d_{\mu}\tilde{\chi}_k\d_{\mu}\tilde{\chi}_{-k}
-k^2\tilde{\chi}_k\tilde{\chi}_{-k}
+\varepsilon_i\omega^{l+2}((ik_z\sqrt{\theta})^l
\tilde{\chi}^*_k\psi_i+h.c.)
\r\r,
\label{mixing}
\end{gather}
here $\psi_i$ are
eigenvectors of the mass matrix for $\psi$ (the second term in
Eq.~(\ref{diagonal})), $\varepsilon_i$ and $\delta_i$ are coefficients of
order one and the extra factor $(k_z\sqrt{\theta})^l$ is due to the
smallness of $\tilde{\chi}_0^l$ (the centrifugal barrier suppression).
Proceeding as before one obtains the following estimate for the
widths of the components of the field $\psi$ of angular momentum $l$,
\begin{gather}
\Gamma\propto {\omega^{4l+3}\over\sqrt{\theta}}\;.
\end{gather}
We conclude that at small $\theta$, all quasi-localized states have
masses of order $\omega/\sqrt{\theta}$, but there is a hierarchy
between their widths: modes with higher angular momenta live longer
on the soliton.

\section{Perturbative analysis}
\label{general}
Before proceeding to other examples of quasi-localization on
noncommutative solitons let us discuss this property in more general
terms.
The purpose of this treatment is twofold. 
On the one hand, it helps to elucidate
the general character of this phenomenon. On the other hand, it provides
a convenient formalism to analyze quasi-localization in more
sophisticated setups than the two-dimensional $U(1)$-Higgs model.

Let us consider a gauge theory\footnote{The gauge group of the
noncommutative theory need not necessarily be $U(1)$; our analysis
applies to noncommutative $U(N)$ gauge theories as well.} (possibly
with some matter fields) 
in space of $2d$ noncommutative coordinates labeled by indices $i,j,k,\dots$
and $p$ commutative coordinates. 
Let us assume that this theory admits a soliton which can be obtained
by the solution generation technique of \cite{Harvey:2000jb}. Namely, the
gauge field is independent of the commutative coordinates, has
vanishing components along commutative coordinates, $A_{\mu}=0, ~\mu =
0,1,\dots,p$, and has the following form in the Fock basis,
\begin{gather}
\label{generalC0}
C_i^{(0)}=S^+a_i^+S\;.
\end{gather}
Here $a_i^+,~~i=1,\dots,d$, are creation operators and $S$ is a
partial isometry operator obeying 
\begin{gather}
SS^+=1\nonumber\\
S^+S=1-P\;,
\label{20}
\end{gather} 
where $P$ is a projector on the $m$-dimensional subspace $V_0$ of the
original Fock space. Let $|\al\rangle$ be the basis in $V_0$,
$\alpha=0,1,\dots,m-1$. The soliton background (\ref{generalC0})
leaves unbroken the subgroup $U(m)$ of the gauge group $U(\infty)$; in
other words, the soliton carries unbroken $U(m)$ gauge theory on its
world-volume. This $U(m)$ may be viewed as the group of unitary
transformations in $V_0$.

Now, let us further  assume that there exists another soliton near
(\ref{generalC0}), whose gauge fields are again independent of the
commuting coordinates. If $U(m)$ is in the Higgs phase on the
world-volume of that soliton, the gauge fields no longer have the form
(\ref{generalC0}). Instead, they are
\begin{gather}
\label{generalC}
C_i=S^+\tilde{a}_i^+S+\Delta_i^+\;.
\end{gather}
where $\Delta_i^+$ obeys
\begin{gather}
S\Delta_i^+S^+=0\;.
\end{gather}
In the vortex example of the previous section, $\tilde{a}$ is given
by Eq.~(\ref{R*}) while $\Delta$ is the sum of (\ref{muvortex})
and (\ref{epsilonvortex}).

We assume that a deviation of the  gauge field 
(\ref{generalC}) from the field obtained by the solution generation
technique is small, so that $\Delta^+$ and $(\tilde{a}_i-a_i)$ may be
considered as small perturbations.
In addition, we assume that the analogue of the 
condition (\ref{zerolimit}) is
satisfied, {\it i.e.}
\begin{equation}
\label{generalzerolimit}
\langle\dots, n_i,\dots|(\tilde{a}_i-a_i)|\dots,k_i,\dots\rangle\to
0,\;{\mathrm as\ }
n_i,\;k_i\to\infty\;,
\end{equation}
where $k_i,n_i$ are occupation numbers for the $i$-th oscillator.

Let us consider a field $f$ in the
adjoint representation of the gauge group of the noncommutative theory.
The field $f$ may be a
scalar as in the previous section, or it may be a component of the
gauge field along the commutative dimensions as will be the case in
the next section. For brevity, in what follows we will drop indices
corresponding to
the Lorentz $SO(1,p)$ group which may be carried by the field
$f$. Then the spectrum for the field $f$ is determined by
its gradient term along noncommutative coordinates, which has the
following form
\begin{equation}
\label{generalkinetic}
{\it M}_{eff}=-{1\over\theta}\Tr\sum_i[C_i,f][C_i^+,f]\;.
\end{equation}
To study the mass spectrum of the field $f$, let us decompose it
in the following way\footnote{It is straightforward to check that
under conditions (\ref{20}) every field $f$ can be written in the
form (\ref{generaldecomposition}) in unique way.},
\begin{equation}
\label{generaldecomposition}
f=\psi_{\al}^{\bt}|\al\rangle\langle\bt|+
|\al\rangle\langle\xi_{\al}|S+S^+|\xi_{\al}\rangle\langle\al|+
S^+\chi S\;,
\end{equation}
where $|\xi_{\al}\rangle$ are arbitrary vectors and $\chi$ is an
arbitrary self-conjugate operator. Notations here are similar to
Eq.~(\ref{D*}). 

To the zeroth order in the deviations
from the field configuration 
obtained by the solution generation technique, the gradient
term (\ref{generalkinetic}) takes the following form,
\begin{equation}
\label{zerokinetic}
{\it M}^{(0)}_{eff}=-{1\over\theta}\sum_i\l\Tr[a_i,\chi][a_i^+,\chi]+
\langle\xi_{\al}|a_i^+a_i+a_ia_i^+|\xi_{\al}\rangle\r\;.
\end{equation}
From this equation one finds the usual structure  of the mass spectrum
for perturbations
around the soliton (\ref{generalC0}):
there is a gapless continuum spectrum parametrized by the operator
field $\chi$, a discrete spectrum of heavy fields, parametrized by a
set of vectors $|\xi_{\al}\rangle$, 
and a finite number of massless excitations localized on
the brane, parametrized by the $m\times m$ matrix
$\psi$. From the $p$-dimensional point of view, each vector
$|\xi_{\al}\rangle$ corresponds to an infinite tower of fields with
masses
\[
m_n^2={2n+1\over\theta}~,~~~n=0,1,\dots
\]
On the other hand, the field $\chi$ has the quadratic action of free
$(1_{time}+p+2d)$-dimensional theory, so it freely propagates in both
commutative and noncommutative dimensions. To the zeroth order, there
is no mixing between $\chi$, $|\xi_{\al}\rangle$ and $\psi$, so the
modes described by $\psi$ and $|\xi_{\al}\rangle$ are strictly
localized on the soliton (\ref{generalC0}). Note also, that if the
adjoint field $f$ is interpreted as components $A_{\mu}$ of the gauge
field along commutative dimensions, the modes $\psi_{\al}^{\beta}$
correspond to the massless $U(m)$ gauge field localized on the soliton.

Let us study perturbation spectrum to higher orders in the deviations
from the soliton (\ref{generalC0}). One
feature is that one should replace operators $a_i$ in
Eq.~(\ref{generalzerolimit}) by operators $\tilde{a}_i$. This change 
introduces mixings between heavy fields $|\xi_{\al}\rangle$ and
modifies the gradient term for the field $\chi$ in the vicinity of the
soliton. However, this modification
will not affect any of the qualitative properties of
the spectrum (\ref{zerokinetic}) due to the asymptotic condition
(\ref{generalzerolimit}). Furthermore, the arguments similar to those
given in the previous section (see Eq.~(\ref{fourier1})) imply
that the corresponding corrections are negligible at small
$(a_i-\tilde{a}_i)$.

In addition, a number of new terms appear. To the linear order in
$\Delta$ these terms are
\begin{equation}
\label{M1}
\begin{split}
M_{eff}^{(1)}&=-{1\over\theta}\sum_i
\l 
2\langle\bt|\Delta_i|\al\rangle\langle
\xi_{\al}|\tilde{a}_i^+|\xi_{\bt}\rangle\right.\\
&\left. -2\langle\xi_{\al}|\tilde{a}_i^+\chi S\Delta_i|\al\rangle+
\langle \xi_{\al}|\chi\tilde{a}_i^+S\Delta_i
|\bt\rangle
+\langle\al|\Delta_i S^+\tilde{a}_i^+\chi|\xi_{\al}\rangle-
2\langle\al|\Delta_i S^+\chi\tilde{a}_i^+|\xi_{\al}\rangle\right.\\
&\left. +\psi_{\al}^{\bt}\langle \bt|\Delta_i S^+ \tilde{a}_i^+|\xi_{\al}\rangle+
\psi^{\al}_{\bt}\langle
\xi_{\al}|\tilde{a}_i^+S\Delta_i|\bt\rangle+h.c.
\r
\end{split}
\end{equation}
This expression is rather lengthy. However, the effects of various
terms are quite clear. The first term introduces additional mixing
between the heavy fields $|\xi_{\al}\rangle$.  The four terms in the
second line are
responsible for mixing between heavy states and continuum spectrum.
As a result of this mixing some of the fields $|\xi_{\al}\rangle$ may
not be strictly localized on the soliton and rather correspond to
quasi-localized
heavy fields. The last two terms introduce mixing between the fields
$\psi_{\al}^{\bt}$ and $|\xi_{\al}\rangle$. The combination of these two
effects in general may lead to the non-trivial mixing between the fields
$\psi_{\al}^{\bt}$ and the states in the continuum spectrum, implying that
the localization of the fields $\psi_{\al}^{\bt}$ is approximate. In fact,
some mixings 
written in Eq.~(\ref{M1}) may vanish in a concrete model.
For instance, as we saw in the previous section, there is 
no mixing, in the vortex background, between the fields 
$\psi_{\al}^{\bt}$ and $|\xi_{\al}\rangle$  to this order.

To the quadratic order in $\Delta$ one has the following mass matrix,
\begin{equation}
\label{M2}
M_{eff}^{(2)}=-{1\over\theta}\sum_i\Tr [\Delta_i,f][\Delta_i^+,f]\;.
\end{equation}
The relevant terms here are as follows.
There are mass terms for the fields $\psi_{\al}^{\bt}$,
\begin{equation}
\label{gpsi2}
-{1\over\theta}\sum_i\Tr [\Delta_i,\psi][\Delta_i^+,\psi]\;.
\end{equation}
Since our analysis applies to the gauge field $A_{\mu}$, these terms
give masses to the gauge bosons of the gauge group $U(m)$ residing on
the soliton; the (quasi-) localized gauge theory on the soliton is in
the Higgs phase. We consider the case of small $\Delta_i$, so the
masses of these gauge fields are small.

Other terms coming from Eq.~(\ref{M2}) are
direct mixings between the fields $\psi_{\al}^{\bt}$ and the field $\chi$
propagating in the bulk,
\begin{equation}
\label{gpsichi}
-{1\over\theta}\sum_i\Tr [\Delta_i,\psi][\Delta_i^+,S^+\chi S]+h.c.\;,
\end{equation}
and new mixings between light fields $\psi_{\al}^{\bt}$ and heavy
fields $|\xi_{\al}\rangle$,
\begin{equation}
\label{gpsixi}
-{1\over\theta}\sum_i\Tr
[\Delta_i,\psi][\Delta_i^+,(S^+|\xi_{\al}\rangle\langle\al|+|\al\rangle
\langle\xi_{\al}|S)]+h.c.\;.
\end{equation}
In the previous section
we saw that the direct mixing between the fields $\psi_{\al}^{\bt}$ and
continuum spectrum (on the main diagonal, $l=0$) and between the heavy
fields $|\xi_{\al}\rangle$ and the light fields $\psi_{\al}^{\bt}$
appeared only from the second order
term (\ref{M2}). 

To summarize, around a soliton close to one
obtained by the solution generation technique, there is a rich pattern
of mixings between would-be localized fields charged under the gauge
group on the soliton and the fields propagating in
the bulk. These mixings lead to quasi-localization of states on the
soliton. Clearly, the necessary condition for these mixings to exist
is that there are non-zero matrix elements of the operator $\Delta$
between subspace of zero modes $V_0$ and its orthogonal complement. In
particular, this condition implies that (part of) the gauge group on
the soliton is in the Higgs phase and that only fields which are
neutral under the unbroken subgroup may mix with bulk modes.

\section{Quasi-localization on noncommutative instantons}
In this section we study quasi-localization of massive gauge fields on
noncommutative instantons. We begin with one-instanton case in $U(2)$
noncommutative gauge theory and make use of the explicit solution
found in Ref.~\cite{Furuuchi:2000dx}. We then generalize to $U(2k)$
noncommutative gauge theory and consider simple $k$-instantons. These
support $U(k)$ gauge theory on their world-volume. Once the
$k$-instanton background is such that this $U(k)$ gauge theory is in
the Higgs phase, the massive gauge bosons become quasi-localized. For
the particular background studied in this section, there is no
hierarchy between the widths of these gauge bosons. However,
hierarchy may be inherent in some $k$-instanton
backgrounds. Indeed, we discuss in Appendix technically more involved
example of a two-instanton solution in $U(2)$ noncommutative gauge
theory, and show that the widths of the quasi-localized gauge bosons
exhibit the hierarchical pattern.

\subsection{One-instanton solution}
\label{oneinstanton}
Let us describe the one-instanton solution in the $U(2)$
noncommutative pure gauge
theory, which was explicitly constructed in Ref.~\cite{Furuuchi:2000dx}. One
considers the $U(2)$
gauge theory in $(1_{time}+p+4)$-dimensional space-time with
four space-like noncommutative dimensions
$z,\bar{z},\zeta,\bar{\zeta}$ and commutative dimensions $y^{\mu}$.
Following Ref.~\cite{Furuuchi:2000dx} we consider the case of anti-self-dual
parameter of noncommutativity, {\it i.e.},
\[
[z,\bar{z}]=-[\zeta,\bar{\zeta}]\equiv\theta>0\;.
\]
In the Fock basis, the action for this theory has the following form,
\begin{equation}
\label{inaction}
S=(2\pi\theta)^2\int d^{p+1}y ~\Tr\left[-{1\over 4}F_{ij}F^{ij}
+{1\over2\theta}
{ D_{\mu}}C_{\bar{\zeta}}{ D^{\mu}}C_{\zeta}+{1\over2\theta}
{ D_{\mu}}C_{\bar{z}}{ D^{\mu}}C_{z}-{1\over
4}F_{\mu\nu}F^{\mu\nu} \right]\;
\end{equation}
where $i=z,\bar{\zeta}$, $C_i$ are
$2\times 2$ matrices whose entries are
operators acting in the Fock
space of two-particle quantum mechanics. 
Components of the field
strength $F_{ij}$ along noncommutative dimensions are
determined by these matrices in the usual way 
\begin{gather}
F_{z\bar{z}}=-{1\over\theta}([C_z,C_{z}^+]+1)\nonumber\\
F_{\zeta\bar{\zeta}}={1\over\theta}([C_{\bar{\zeta}},C_{\bar{\zeta}}^+]+1)
\nonumber\\
F_{\zeta z}=-{1\over\theta}[C_{\bar{\zeta}}^+,C_{z}]\nonumber\\
F_{\zeta\bar{z}}=-{1\over\theta}[C_{\bar{\zeta}}^+,C_{z}^+]\;.\nonumber
\end{gather}
The covariant derivative of the field $C_{z}$ is 
\begin{gather}
D_{\mu}C_{z}=\d_{\mu}C_{z}-i[A_{\mu},C_{z}]
\end{gather}
and the same for the field $C_{\bar{\zeta}}$. Recall that
$C_{\bar{z}}=C_z^+$, $C_{\zeta}=C_{\bar{\zeta}}^+$, and that the
vacuum is $C_z=a_z^+\cdot {\mathbf 1}$,
$C_{\bar{\zeta}}=a_{\zeta}^+\cdot {\mathbf 1}$, where ${\mathbf 1}$ is
$2\times 2$ unit matrix (which we will not write explicitly in what
follows). 

The instanton solution is independent of commutative coordinates, has
$\mu$-components of the gauge field equal to zero and
has anti-self-dual field strength. 
As in the commutative case, the Pontryagin
index of the $m$-instanton solution,
\begin{equation}
\label{pont}
N_P=-{\theta^2\over 8}\epsilon^{ijkl}\Tr F_{ij}F_{kl}\;,
\end{equation}
is equal to $m$. In Eq.~(\ref{pont}) 
the trace is evaluated over both the $U(2)$ indices and the
Fock space. 

A powerful tool for describing the moduli space of
$m$-instanton solutions and obtaining explicit formulas for instanton
fields is the noncommutative version~\cite{Nekrasov:1998ss} of the ADHM 
construction~\cite{Atiyah:ri}. Explicit construction of the one-instanton
solution in the noncommutative $U(2)$ theory involves the $4\times 2$
matrix $\Psi$~\cite{Furuuchi:2000dx} which can be written in the following form,
\begin{gather}
\label{psi}
\Psi=
\l\begin{array}{c}
S^+{\rho\over\sqrt{N+1+\rho^2}}S+1-S^+S\\ \\
{\sqrt{N+1}\over\sqrt{N+1+\rho^2}}S
\end{array}\r
\end{gather}
where entries are $2\times 2$ matrices, the matrix $U$ is 
\begin{equation}
\label{U}
S={1\over \sqrt{N+1}}\l
\begin{array}{cc}
a_{\zeta}^+ & a_z\\
-a_z^+ & a_{\zeta}
\end{array}\r\;,
\end{equation}
and $N$ is the occupation number operator,
$N=a_z^+a_z+a_{\zeta}^+a_{\zeta}$. The real
parameter $\rho$ is natural 
to interpret as the size of the instanton in units of $\sqrt{\theta}$.
The gauge fields $C_z(\rho)$, $C_{\bar{\zeta}}(\rho)$ 
of the instanton of size $\rho$ are
\begin{gather}
C_z(\rho)=\Psi^+a_z^+\Psi\nonumber\\
C_{\bar{\zeta}}(\rho)=\Psi^+a_{\zeta}^+\Psi\;.
\label{Cs}
\end{gather}
It is straightforward to check that the field strength obtained
from $C_z(\rho)$, $C_{\bar{\zeta}}(\rho)$ is anti-self-dual and
has unit Pontryagin index~\cite{Furuuchi:2000dx}.

The operator $S$ is a partial isometry
operator, {\it i.e.}
\begin{gather}
SS^+=1\nonumber\\
S^+S=1-P_0\;,
\end{gather}
where $P_0$ is the projector on the state
\begin{equation}
\label{alone}
|\alpha\rangle=
\l
\begin{array}{c}
0\\
|0\rangle
\end{array}
\r\;.
\end{equation}
The instanton of zero size, $\rho=0$, is non-singular and may be obtained
from vacuum
by the solution generation technique,
\begin{gather}
C_z(\rho=0)=S^+a_z^+S\nonumber\\
C_{\bar{\zeta}}(\rho=0)=S^+a_{\zeta}^+S
\end{gather}
Consequently, the instanton of zero size
supports unbroken $U(1)$ gauge group on its world-volume. 

\subsection{Quasi-localization on single instanton}
Let us study the spectrum of the components $A_{\mu}$ along the
commutative directions, in the background of the instanton of small
but non-vanishing size $\rho$. We apply the technique of
Sect.~\ref{general}, with $A_{\mu}$ playing the role of the field $f$.

In the
notation of Sect.\ref{general} the operator $\psi$ which describes
the would-be zero mode of $A_{\mu}$ is (in what follows we drop
the index $\mu$ everywhere)
\begin{gather}
\label{azero}
\psi=
\psi_0(y)\l\begin{array}{cc}
0 & 0\\
0 & |0\rangle\langle 0|
\end{array}\r\;.
\end{gather}
Now, it is straightforward to present the instanton of non-zero size
$\rho$ in the form (\ref{generalC}) with
\begin{gather}
\tilde{a}_z=
{1\over\sqrt{N+1+\rho^2}}
\l\sqrt{N+1}\;a_z\sqrt{N+1}+\rho^2S^+a_zS\r 
{1\over\sqrt{N+1+\rho^2}}
\label{atilde}\\
\Delta_z={\rho\over\sqrt{1+\rho^2}}|\alpha\rangle\langle v_0|\;,
\label{rhoinst}
\end{gather}
where
\begin{gather}
|v_0\rangle=\l
\begin{array}{c}
0\\
a_z^+|0\rangle
\end{array}
\r\;.
\label{v0}
\end{gather}
The expressions for the field $C_{\bar{\zeta}}$ are obtained from
Eqs.~(\ref{atilde}) -- (\ref{v0}) by substituting $a_{\zeta}$ for
$a_z$. Plugging this
form of $\Delta$ into the general expression (\ref{M1}) determining
the perturbation spectrum about the instanton to the linear order in $\Delta$,
one obtains, that the only non-zero term is 
\[
M_{\chi\xi}={2\over\theta}\sum_{i=z,\zeta} \langle 0|\Delta_i
S^+\chi \tilde{a}_i^+|\xi_0\rangle\;. 
\]
This term is
responsible for the mixing of the heavy field $|\xi_0\rangle$ with the
continuum. 

This situation is quite similar to what we have found for
one-vortex solution in Sect.~\ref{vortex}, so let us consider the
mass matrix to the second order in $\Delta$, Eq.~(\ref{M2}).
Among other terms we obtain the mass term for the would-be zero mode
$\psi_0(y)$
(cf. Eq.~(\ref{gpsi2}))
\begin{equation}
\label{psi2}
M_{\psi\psi}={2\rho^2\over \theta (1+\rho^2)}\psi_0^2
\end{equation}
and direct mixing between the field $\psi_0(y)$ and continuum
(cf. Eq.~(\ref{gpsichi}))
\begin{equation}
\label{psichi}
M_{\psi\chi}=
-{2\rho^2\over \theta (1+\rho^2)}\psi_0
\l\langle \alpha|\chi|\alpha\rangle +\langle u|\chi|u\rangle\r\;,
\end{equation}
where
\begin{gather}
|u\rangle=
\l
\begin{array}{c}
|0\rangle\\
0
\end{array}
\r\;.
\label{ucolumn}
\end{gather}
Since there is no first order mixing between the light mode $\psi$ and heavy modes
$|\xi_0\rangle$, mixing involving the heavy states $|\xi_0\rangle$ is
negligible at small $\rho$ and low energies.

The rest of the analysis is the same as
in the case of vortex.
The effect of the two terms (\ref{psi2}), (\ref{psichi}) is that the field
$\psi_{0}$ describes a quasi-localized massive vector field on the
soliton world-volume, whose mass and width (at small $\rho$) are
\begin{gather}
\label{minst}
m_0^2={2\rho^2\over\theta}\\
\label{Ginst}
\Gamma= {\pi\sqrt{2}\rho^5\over\sqrt{\theta}}\;,
\end{gather}
The extra factor $\rho^2$ in the expression for the width, as compared to the
case of vortex (cf. Eq.~(\ref{Gvortex}), is due to the fact that there
are four, rather than two, transverse dimensions in the case of instanton.

\subsection{Multi-instanton case}
In this subsection we discuss quasi-localization on a
multi-instanton. Similar to the case of vortex, the main difference
from the one-instanton case is that an $m$-instanton
solution supports a {\it non-Abelian} gauge group $U(m)$ on its
world-volume. 
If all $m$ instantons sit on top of each other, this $U(m)$ is unbroken.
One way to break this gauge group spontaneously is to move to a
general point in the Coulomb branch where positions of instantons in
the noncommutative hyperplane are not coincident. This splitting
leaves unbroken a subgroup $[U(1)]^m$ of $U(m)$. The massive gauge
bosons on the soliton all carry non-zero charges corresponding to some
of the $U(1)$ factors, so they remain strictly localized.

We consider instead the case in which instantons have non-zero
sizes. This corresponds to the Higgs branch of the instanton moduli
space.
A simple solution of this kind may be obtained in $U(2k)$ gauge
theory by making use of the one-instanton solution considered in the
previous subsection. For the sake of simplicity, let us consider
two-instanton
solution in the $U(4)$ gauge theory; a generalization to $k$-instanton
solution in $U(2k)$ gauge theory is straightforward. The gauge field 
of a simple anti-self-dual solution
describing two instantons of sizes $\rho_1$ and $\rho_2$ sitting on
top of each other has the following block-diagonal form,
\begin{equation}
\label{simpletwo}
C_z(\rho_1,\rho_2)=\l
\begin{array}{cc}
C_z(\rho_1) & 0\\
0 & C_z(\rho_2)
\end{array}\r\;,
\end{equation}
and analogously for $C_{\bar{\zeta}}(\rho_1,\rho_2)$.
Here $C_z(\rho_i)$ are $2\times 2$ matrices describing one-instanton
solution in the $U(2)$ gauge theory, see Eq.~(\ref{Cs}). 
Clearly, the field strength corresponding to Eq.~(\ref{simpletwo}) is
anti-self-dual and has Pontryagin index equal to two.

When both instantons have zero sizes, $\rho_1=\rho_2=0$, this
solution may be obtained from vacuum by the solution generation
technique with the partial isometry operator
\begin{equation}
\label{S2}
S_2=\l
\begin{array}{cc}
S & 0\\
0 & S
\end{array}\r\;,
\end{equation}
where $S$ is given by Eq.~(\ref{U}). In this case there is an unbroken
$U(2)$ gauge group, which corresponds to unitary transformations in
the two-dimensional subspace $V_0$ of the Fock space, whose basis
vectors are
\begin{equation}
|\al_1\rangle=\l
\begin{array}{c}
|\al\rangle\\
0
\end{array}\r,\;\;
|\al_2\rangle=\l
\begin{array}{c}
0\\
|\al\rangle
\end{array}\r,
\end{equation}
where the two-column $|\al\rangle$ is given by
Eq.~(\ref{alone}). The four real zero modes $\psi$ of the field
$A_{\mu}$, corresponding to this gauge group, can be organized as
follows,
\begin{equation}
\label{psimult}
\psi=
\l
\begin{array}{cc}
\psi^1_1(y)& \psi_1^2(y)\\
\psi_2^1(y)& \psi_2^2(y)
\end{array}
\r\otimes |\al\rangle\langle\al |\;,
\end{equation}
where 
\[
\psi_1^2=\psi_2^{1*}\;.
\]
and $\psi_1^1$, $\psi_2^2$ are real.

When
both instantons have small but non-zero sizes, $\rho_1\neq 0$,
$\rho_2\neq 0$, the
$U(2)$ gauge group is completely Higgsed, and all its gauge fields
become massive. Their mass matrix is obtained by plugging the fields
(\ref{psimult}) and (\ref{simpletwo}) 
into Eq.~(\ref{gpsi2}) with the result
\begin{gather}
M_{\psi\psi}={2\over\theta}\l {\rho_1^2\over 1+\rho_1^2}(\psi_1^1)^2+
{\rho_2^2\over 1+\rho_2^2}(\psi_2^2)^2 +\l{\rho_1^2\over 1+\rho_1^2}+
{\rho_2^2\over 1+\rho_2^2}\r\psi_1^2\psi_2^1\r\;
\label{psi2multi}
\end{gather} 
Similarly to the one-instanton case, there is no mixing between the fields
$\psi_{\al}^{\bt}$ and heavy charged fields $|\xi_{\al}\rangle$ to the linear
order and the leading contributions to the widths of
$\psi_{\al}^{\bt}$ come from direct mixing with the fields from
the continuum, Eq.~(\ref{gpsichi}). The diagonal components
$\psi_1^1$ and $\psi_2^2$ mix with the corresponding diagonal
components $\chi_1^1$ and $\chi_2^2$. Each of these mixings has
precisely the same form as in the one-instanton case,
Eq.~(\ref{psichi}), leading to the widths
\begin{equation}
\Gamma_{11} = {\pi\sqrt{2}\rho_1^5\over\sqrt{\theta}},~\;
\Gamma_{22} = {\pi\sqrt{2}\rho_2^5\over\sqrt{\theta}}\;.
\end{equation}
The off-diagonal component $\psi_1^2$ mixes with the off-diagonal
 component $\chi_2^1$ of the field $\chi$. This mixing has the
 following form
\begin{equation}
\label{2psichi}
M_{\psi\chi}^{off-diag}=
-{2\rho_1\rho_2\over \theta \sqrt{(1+\rho_1^2)(1+\rho_2^2)}}\psi_1^2
\l\langle \alpha|\chi_2^1|\alpha\rangle +\langle
u|\chi_2^1|u\rangle\r+h.c.
\;,
\end{equation}
where $|u\rangle$ is still given by Eq.~(\ref{ucolumn}).
This mixing leads to the width of $\psi_1^2$,
\begin{equation}
\Gamma_{12} = {\pi\sqrt{2}(\rho_1\rho_2)^2(\rho_1^2+\rho_2^2)^{1/2}
\over\sqrt{\theta}}.\;
\end{equation}
We see that when one of the instantons has 
zero size, one of the diagonal gauge bosons is massless, which is related
to the fact that
the $U(2)$ gauge group on the instanton world-volume
 is broken down to its $U(1)$ subgroup. In
this case the off-diagonal component of the vector field is
massive, see Eq.~(\ref{psi2multi}). However, its width is equal to
zero. This result is in 
complete agreement with our general reasoning in
section~\ref{general}, as this component describes massive vector
field charged under the unbroken $U(1)$ gauge group in this case.

If $\rho_1$ and $\rho_2$ are of the same order,
$\rho_1\sim\rho_2\sim\rho$, there is no hierarchy between the masses
and widths of the quasi-localized gauge bosons: all three masses are
of order $\rho/\sqrt{\theta}$, and all three widths are smaller by a
factor $\rho^4$. The hierarchy of widths 
may be inherent in other noncommutative instanton backgrounds: in
Appendix we present a two-instanton solution in whose background
the widths of quasi-localized gauge bosons exhibit the hierarchy
similar to one appearing in the multi-vortex case.

\section*{Acknowledgments}
The authors are indebted to F.L.~Bezrukov for useful discussions.
This work has been supported in part by RFBR grant 99-02-18410, CPG
and SSLSS grant 00-1596626, CRDF grant (award RP1-2103), Swiss Science
Foundation grant 7SUPJ062239. The work of S.S. was supported in part
also under RFBR grant 01-02-06034.
\appendix
\section{Quasi-localization in two-instanton background in noncommutative
$U(2)$ gauge theory}
To obtain an explicit
two-instanton solution in noncommutative $U(2)$ gauge theory, we make use
of the ADHM construction. We begin with the following solution of the
ADHM equations,
\begin{gather}
\begin{array}{cc}
B_1=\l
\begin{array}{cc}
0&0\\
\rho&0
\end{array}\r
&
B_2=\l
\begin{array}{cc}
0&\rho\\
0&0
\end{array}\r
\end{array}
\label{Bs}
\\
\begin{array}{cc}
I=\l
\begin{array}{cc}
\rho&0\\
0&\rho
\end{array}\r
&
J=\l
\begin{array}{cc}
\rho&0\\
0&-\rho
\end{array}\r
\end{array}
\label{IJ}
\end{gather}
With these ADHM data, one should find two ``zero modes'' $\Psi^a$
($a=1,2$) which are solutions
of the following equation
\begin{equation}
\label{Dz}
{\cal D}\Psi^a=0\;,
\end{equation}
where
\[
{\cal D}=\l
\begin{array}{c}
\tau\\
\sigma^+
\end{array}\r
\]
is a $4\times 6$ matrix whose $2\times 6$ entries are
\[
\tau=
\l
\begin{array}{ccc}
B_2-a_z,& B_1-a_{\zeta}^+,& I
\end{array}
\r
\]
and 
\[
\sigma^+=
\l
\begin{array}{ccc}
-(B_1^+-a_{\zeta}),& B_2^+-a_z^+,& J^+
\end{array}
\r\;.
\]
The two linear independent solutions to Eq.~(\ref{Dz}) are
\begin{equation}
\label{firstpsi}
\Psi^1=
\l
\begin{array}{c}
0\\
\sqrt{2}\rho a_-^+a_+^+\\
2\rho^2a_+^+\\
\sqrt{2}\rho\,(2\rho^2+N_-+1)\\
\sqrt{2}\rho(a_+^+)^2\\
(2\rho^2+N_++N_-+1)\,a_-^+
\end{array}\r \cdot{\cal N}_1
\end{equation}
and
\begin{equation}
\label{secondpsi}
\Psi^2=
\l
\begin{array}{c}
\sqrt{2}\rho\, (2\rho^2+N_-)\\
2\rho^2a_+\\
-\sqrt{2}\rho a_+a_-\\
0\\
-a_-(2\rho^2+N_++N_-)\\
\sqrt{2}\rho a_+^2
\end{array}\r \cdot{\cal N}_2
\end{equation}
where
\begin{gather*}
a_{\pm}={a_{\zeta}\pm a_z\over\sqrt{2}}\;,\\
N_{\pm}=a^+_{\pm}a_{\pm}\;,
\end{gather*}
${\cal N}_{1,2}$ are normalization factors and $\rho$ is a real
parameter which may be interpreted as the instanton size. We will need
the expression for ${\cal N}_2$ only,
\[
\begin{split}
{\cal N}_2=\left[ 2\rho^2(2\rho^2+N_-)^2+(2\rho^2)^2 N_+
+2\rho^2N_+N_- +(N_++N_-+2\rho^2)^2N_-\right.\\
\left.+2\rho^2N_+(N_+-1)\right]^{-{1\over2}}
\end{split}
\]
Let us first consider the limit $\rho\to 0$. In this limit the solutions
(\ref{firstpsi}) and (\ref{secondpsi}) take the following form,
\begin{equation}
\label{firstpsi0}
\Psi_{(0)}^1=
\l
\begin{array}{c}
0\\
0\\
0\\
0\\
0\\
a_-^+{1\over\sqrt{N_-+1}}
\end{array}\r
\end{equation}
and
\begin{equation}
\label{secondpsi0}
\Psi_{(0)}^2=
\l
\begin{array}{c}
|0_+,0_-\rangle\langle 0_+,0_-|\\
|0_+,0_-\rangle\langle 1_+,0_-|\\
0\\
0\\
-{1\over\sqrt{N_-+1}}a_-\\
{1\over\sqrt{(N_++1)(N_++2)}}a_+^2\otimes
|0_-\rangle\langle 0_-| 
\end{array}\r
\end{equation}
where $|1_+,0_-\rangle=a_+^+|0_+,0_-\rangle$, etc.
The corresponding gauge fields are
\begin{gather}
C_z^{(0)ab}=\Psi_{(0)}^{a\dagger }a_z^+\Psi_{(0)}^b\;,\\
C_{\bar{\zeta}}^{(0)ab}=\Psi_{(0)}^{a\dagger}a_{\zeta}^+\Psi_{(0)}^b\;.
\end{gather}
It is straightforward to check that the gauge fields
$C^{(0)}_z$, $C^{(0)}_{\bar{\zeta}}$ may be obtained from vacuum by
applying the
solution generation technique with the following partial isometry
operator
\begin{gather}
\label{SSS}
S=\l
\begin{array}{cc}
0& -{1\over\sqrt{N_-+1}}a_-\\
a_-^+{1\over\sqrt{N_-+1}}& 
{1\over\sqrt{(N_++1)(N_++2)}}a_+^2\otimes
|0_-\rangle\langle 0_-| 
\end{array}\r
\end{gather}
The subspace $V_0$ of zero vectors of this operator is
two-dimensional, the basis vectors being
\[
\begin{array}{cc}
|\al_1\rangle=\l
\begin{array}{c}
0\\
|0_+,0_-\rangle
\end{array}\r
&
|\al_2\rangle=\l
\begin{array}{c}
0\\
|1_+,0_-\rangle
\end{array}\r
\end{array}
\]
Consequently, at $\rho=0$ there is unbroken $U(2)$ gauge group on the
world-volume of the two-instanton. To consider the case of
non-zero $\rho$ one should note that the vectors (\ref{firstpsi}) and
(\ref{secondpsi}) are not orthogonal at $\rho\neq 0$,
\[
\Psi^{1\dagger}\Psi^2={-\sqrt{2}\rho\over
(N_-+1)(N_++N_-+2)(N_++N_-+3)} a_+^2a_-+O(\rho^2)\;.
\]
To make them orthogonal, we replace $\Psi^1$ with the following linear
combination 
\[
\tilde{\Psi}^1=\l\Psi^1-\Psi^2(\Psi^{2\dagger}\Psi^1)\r\tilde{\cal N}_1\;,
\]
where $\tilde{\cal N}_1$ is a new normalization factor.

As follows from Eqs.~(\ref{M1}), (\ref{gpsi2}) and
(\ref{gpsixi}), to calculate the mass matrix of light quasi-localized
modes $\psi_{\al}^{\bt}$ and their mixings with the continuum modes
$\chi$ one does not need the explicit form of the gauge fields $C_i$. It
is sufficient to calculate the action of these operators in the
space\footnote{Note that the operators $\Delta_i^+$ entering   
Eqs.~(\ref{M1}), (\ref{gpsi2}) and
(\ref{gpsixi}) act in $V_0$ in the same way as the operators $C_i$.}
$V_0$. 
It is straightforward to perform this calculation to the leading order
in $\rho$, with the following result,
\begin{gather}
\label{acton}
\begin{array}{cc}
C_+|\al_1\rangle=
\l
\begin{array}{c}
0\\
\sqrt{2}\rho |1_+,0_-\rangle
\end{array}
\r&
C_-|\al_1\rangle=
\l
\begin{array}{c}
0\\
\sqrt{2}\rho |0_+,1_-\rangle
\end{array}\r\\
C_+|\al_2\rangle=
\l
\begin{array}{c}
0\\
\sqrt{2}\rho |2_+,0_-\rangle
\end{array}
\r&
C_-|\al_2\rangle=
\l
\begin{array}{c}
0\\
2\rho^2 |1_+,1_-\rangle
\end{array}\r\\
C^+_+|\al_2\rangle=
\l
\begin{array}{c}
0\\
\sqrt{2}\rho |0_+,0_-\rangle
\end{array}
\r&
\end{array}\;,
\end{gather}
All the rest are zero. 

With these expressions at hand it is straightforward to calculate all masses
and mixings of interest. Making use of Eq.~(\ref{gpsi2}) one obtains the
following mass matrix
for fields $\psi_{\al}^{\bt}$,
\[
M_{\psi\psi}={4\rho^2\over 
\theta}\left[(\psi_1^1)^2+(\psi_2^2)^2-\psi_1^1\psi_2^2+
2\psi_1^2\psi_2^1)\right]
\]
Similarly to all cases considered previously, it is straightforward to check
that there is no mixing between the fields $\psi_{\alpha}^{\bt}$ and
the heavy
fields $|\xi_{\al}\rangle$ to the linear
order in $\rho$. Consequently, direct  mixing between the fields 
$\psi_{\alpha}^{\bt}$ and continuum states $\chi$, coming from the term
given by 
Eq.~(\ref{gpsichi}), again gives the leading order contribution 
to the widths of $\psi_{\alpha}^{\bt}$. By making use of 
Eqs.~(\ref{acton}) and explicit form of the partial isometry operator 
(\ref{SSS}) it is straightforward to check that the diagonal components
$\psi_1^1$ and $\psi_2^2$ mix with the continuum at the order
$\rho^2$, while mixing of the off-diagonal components
$\psi_1^2=\psi_2^{1*}$ occurs through $O(\rho^2)$ term in
Eq.~(\ref{acton}). As a result, the latter fields mix with the
continuum only at the  order $\rho^3$. Consequently, similarly to the
multi-vortex case we obtain a hierarchical pattern of widths, namely
\[
\Gamma_d\propto {\rho^5\over\sqrt{\theta}}
\]
for the diagonal components and
\[
\Gamma_o\propto {\rho^7\over\sqrt{\theta}}
\]
for the off-diagonal components.


\begin{thebibliography}{99}
\bibitem{Charmousis:1999rg}
C.~Charmousis, R.~Gregory and V.~A.~Rubakov,
Phys.\ Rev.\ D {\bf 62} (2000) 067505
[arXiv:hep-th/9912160];\\
R.~Gregory, V.~A.~Rubakov and S.~M.~Sibiryakov,
Phys.\ Rev.\ Lett.\  {\bf 84} (2000) 5928
[arXiv:hep-th/0002072];\\
C.~Csaki, J.~Erlich and T.~J.~Hollowood,
Phys.\ Rev.\ Lett.\  {\bf 84} (2000) 5932
[arXiv:hep-th/0002161].\\
G.~R.~Dvali, G.~Gabadadze and M.~Porrati,
Phys.\ Lett.\ B {\bf 484} (2000) 112
[arXiv:hep-th/0002190].
\bibitem{Kogan:1999wc}
I.~I.~Kogan, S.~Mouslopoulos, A.~Papazoglou, G.~G.~Ross and J.~Santiago,
Nucl.\ Phys.\ B {\bf 584} (2000) 313
[arXiv:hep-ph/9912552];\\
I.~I.~Kogan, S.~Mouslopoulos, A.~Papazoglou and G.~G.~Ross,
Nucl.\ Phys.\ B {\bf 595} (2001) 225
[arXiv:hep-th/0006030].

\bibitem{Dvali:2000hr}
G.~R.~Dvali, G.~Gabadadze and M.~Porrati,
Phys.\ Lett.\ B {\bf 485} (2000) 208
[arXiv:hep-th/0005016].

\bibitem{Dvali:2000rx}
G.~R.~Dvali, G.~Gabadadze and M.~A.~Shifman,
Phys.\ Lett.\ B {\bf 497} (2001) 271
[arXiv:hep-th/0010071].

\bibitem{Dubovsky:2000am}
S.~L.~Dubovsky, V.~A.~Rubakov and P.~G.~Tinyakov,
Phys.\ Rev.\ D {\bf 62} (2000) 105011
[arXiv:hep-th/0006046].

\bibitem{Dubovsky:2001fj}
S.~L.~Dubovsky,
``Tunneling into extra dimension and 
high-energy violation of Lorentz  invariance,''
arXiv:hep-th/0103205.

\bibitem{Rubakov:2001kp}
V.~A.~Rubakov,
Uspekhi Fiz. Nauk, {\bf 171} (2001) 913,
[arXiv:hep-ph/0104152].

\bibitem{Nekrasov:2000ih}
N.~A.~Nekrasov,
``Trieste lectures on solitons in noncommutative gauge theories,''
arXiv:hep-th/0011095.

\bibitem{Harvey:2001yn}
J.~A.~Harvey,
``Komaba lectures on noncommutative solitons and D-branes,''
arXiv:hep-th/0102076.
\bibitem{Konechny:2001wz}
A.~Konechny and A.~Schwarz,
``Introduction to M(atrix) theory and noncommutative geometry'', Part I
arXiv:hep-th/0012145;
 Part II,
arXiv:hep-th/0107251.



\bibitem{Connes:1997cr}
A.~Connes, M.~R.~Douglas and A.~Schwarz,
JHEP {\bf 9802} (1998) 003
[arXiv:hep-th/9711162];\\
M.~R.~Douglas and C.~M.~Hull,
JHEP {\bf 9802} (1998) 008
[arXiv:hep-th/9711165];\\
V.~Schomerus,
JHEP {\bf 9906} (1999) 030
[arXiv:hep-th/9903205];\\
N.~Seiberg and E.~Witten,
JHEP {\bf 9909} (1999) 032
[arXiv:hep-th/9908142].

\bibitem{Pilo:2000nc}
L.~Pilo and A.~Riotto,
JHEP {\bf 0103} (2001) 015
[arXiv:hep-ph/0012174];\\
\bibitem{Harvey:2000jt}
J.~A.~Harvey, P.~Kraus, F.~Larsen and E.~J.~Martinec,
JHEP {\bf 0007} (2000) 042
[arXiv:hep-th/0005031].

\bibitem{Gopakumar:2000rw}
R.~Gopakumar, S.~Minwalla and A.~Strominger,
JHEP {\bf 0104} (2001) 018
[arXiv:hep-th/0007226].

\bibitem{Douglas:1995bn}
M.~R.~Douglas,
``Branes within branes,''
arXiv:hep-th/9512077.

\bibitem{Polychronakos:2000zm}
A.~P.~Polychronakos,
Phys.\ Lett.\ B {\bf 495} (2000) 407
[arXiv:hep-th/0007043].

\bibitem{Jatkar:2000ei}
D.~P.~Jatkar, G.~Mandal and S.~R.~Wadia,
JHEP {\bf 0009} (2000) 018
[arXiv:hep-th/0007078].

\bibitem{Bak:2000ac}
D.~Bak,
Phys.\ Lett.\ B {\bf 495} (2000) 251
[arXiv:hep-th/0008204].

\bibitem{Bak:2000im}
D.~Bak, K.~Lee and J.~H.~Park,
Phys.\ Rev.\ D {\bf 63} (2001) 125010
[arXiv:hep-th/0011099].

\bibitem{Nekrasov:1998ss}
N.~Nekrasov and A.~Schwarz,
Commun.\ Math.\ Phys.\  {\bf 198} (1998) 689
[arXiv:hep-th/9802068].

\bibitem{Aganagic:2000mh}
M.~Aganagic, R.~Gopakumar, S.~Minwalla and A.~Strominger,
JHEP {\bf 0104} (2001) 001
[arXiv:hep-th/0009142].

\bibitem{Furuuchi:2000vc}
K.~Furuuchi,
``Topological charge of U(1) instantons on noncommutative R**4,''
arXiv:hep-th/0010006.

\bibitem{Harvey:2000jb}
J.~A.~Harvey, P.~Kraus and F.~Larsen,
JHEP {\bf 0012} (2000) 024
[arXiv:hep-th/0010060].

\bibitem{Furuuchi:2000dx}
K.~Furuuchi,
JHEP {\bf 0103} (2001) 033
[arXiv:hep-th/0010119].

\bibitem{Dubovsky:2000av}
S.~L.~Dubovsky, V.~A.~Rubakov and P.~G.~Tinyakov,
JHEP {\bf 0008} (2000) 041
[arXiv:hep-ph/0007179].

\bibitem{Hamanaka} M.~Hamanaka,
{\it ADHM/Nahm construction of localized solitons
in noncommutative gauge  theories},
hep-th/0109070.       

\bibitem{Atiyah:ri}
M.~F.~Atiyah, N.~J.~Hitchin, V.~G.~Drinfeld and Y.~I.~Manin,
Phys.\ Lett.\ A {\bf 65}, 185 (1978).



\end{thebibliography}
\end{document}